\documentclass[aps,prb,showpacs,twocolumn,superscriptaddress]{revtex4}
\usepackage[english]{babel}
\usepackage{graphicx}
\usepackage{color}
\usepackage[utf8]{inputenc}
\usepackage{pstricks,pst-grad,color}
\usepackage{graphicx,pifont,amssymb}
\usepackage{amssymb}

\begin{document}

%\preprint{}

\title{Spin Density Waves in the Hubbard model - A DMFT approach}



\author{Robert Peters}
\email[]{peters@scphys.kyoto-u.ac.jp}
\affiliation{Department of Physics, Kyoto University, Kyoto 606-8502, Japan}
\affiliation{Computational Condensed Matter Physics Laboratory, RIKEN, Wako, Saitama 351-0198, Japan}

\author{Norio Kawakami}
\affiliation{Department of Physics, Kyoto University, Kyoto 606-8502,  Japan}

\date{\today}

%%%%%%%%%%%%%%%%%%%%%%%%%%%%%%%%%%%%%%%%%%%%%%%%%%%%%%%5
\begin{abstract}
We analyze spin density waves (SDWs) in the Hubbard model on a square
lattice within the
framework of inhomogeneous dynamical mean field theory (iDMFT). Doping
the half-filled Hubbard model results in a change of the
antiferromagnetic N\'eel state, 
which exists exactly at half filling, to a phase of
incommensurate SDWs. Previous studies of this phase mainly rely on
static mean field calculations.
In this paper, we will use large-scale  
iDMFT calculations to study properties of SDWs in the Hubbard model. 
A great advantage of iDMFT over
static mean field  approaches is the inclusion of local screening
effects and the easy access to
dynamical correlation functions. Furthermore, this technique is not
restricted to the Hubbard model, but can be easily used to study
incommensurate phases in various strongly correlated materials.
\end{abstract}
%%%%%%%%%%%%%%%%%%%%%%%%%%%%%%%%%%%%%%%%%%%%%%%%%%%%%%%%%%%%

%\keywords{}

\maketitle

%%%%%%%%%%%%%%%%%%%%%%%%%%%%%%%%%%%%%%%%%%%%%%%%%%%%%%%%%%%%%%%%
\section{Introduction}
Strongly correlated materials have been the focus of interest for over
half a century, because of their intriguing properties such as
metal-insulator transitions, magnetism and high-temperature 
superconductivity, which cannot be observed in
weakly-interacting systems.
A prototype model for theoretically describing strongly correlated materials
is the one-band Hubbard model,\cite{hubbard1963,kanamori1963,gutzwiller1963}
\begin{equation}
H=\sum_{ij,\sigma}t_{ij}c_{i\sigma}^\dagger c_{j\sigma}+U\sum_in_{i\downarrow}n_{i\uparrow},\label{hubbard}
\end{equation}
where the first term corresponds to the kinetic energy and the second term to a
local density-density interaction. The operator $c_{i\sigma}^\dagger$
creates an electron on lattice site $i$ in spin direction $\sigma$,
and the operator $n_{i \sigma}=c^\dagger_{i \sigma} c_{i \sigma}$
corresponds to the electron density at site $i$. The interaction is taken to be
repulsive, $U>0$, throughout this paper.

The physics of the Hubbard model with repulsive interaction is
determined by the competition between the local density-density
interaction and the non-local kinetic energy.
This competition is the cause for the Mott-metal-insulator transition,
a well known phenomenon observed in the Hubbard model.\cite{imada1998}
The half-filled 
Hubbard model undergoes a transition from a
metal to a Mott insulator, where due to the repulsive interaction, electrons
become localized.
Besides the metal-insulator transition,
long-range ordered phases have also been extensively studied in the Hubbard
model. For a bipartite lattice and large enough dimensions, $d>1$, the
ground state of the Hubbard model at half filling is an
antiferromagnetic N\'eel state;
each lattice site is occupied with one electron in average, and the spin
polarization alternates between neighboring lattice sites.
Besides this antiferromagnetic phase at half filling, one can also
observe different ordered phases like ferromagnetism or
superconductivity in the Hubbard model, depending on the lattice
structure and
system parameters. 

There are a variety of
analytical and numerical techniques to theoretically analyze the
Hubbard model. A particularly successful  
technique, which is able to directly study the properties of
strongly correlated models, is the dynamical mean field
theory (DMFT).\cite{Metzner1989,Pruschke1995,Georges1996} DMFT maps
the lattice model onto a quantum impurity model, which must be
solved self-consistently. Non-local
terms in the self-energy are thereby neglected, which becomes exact for
infinite dimensional lattices. Although DMFT is an approximation for
real materials, it has provided many insights into fundamental
properties of strongly correlated materials. Furthermore, there are
ways to incorporate the momentum dependence into the self-energy, which
are known as cluster DMFT or dynamical cluster approximation.\cite{Maier2006}

Although long-range ordered phases have been analyzed by
DMFT since the introduction of the method, previous works have mainly
focused on commensurate phases like the antiferromagnetic N\'eel
state, ferromagnetism, etc. 
In order to analyze the antiferromagnetic N\'eel state within DMFT,\cite{Jarrell1992,Freericks_1995,dongen1995,dongen1996,zitzler2002,Peters2009a,Peters2009b}
the lattice is divided into two-site clusters, and
momentum-independent self-energies are calculated separately for each
sublattice. This method is known as the two-sublattice method.

The antiferromagnetic state with electron density close to, but away
from unity has little been analyzed within DMFT. One
approach to perform DMFT 
calculations for such incommensurate states has been to incorporate a fixed
rotation angle of the spin direction into the DMFT equations.\cite{Fleck_1998,Fleck_1999} However,
if the assumed rotation angle does not correspond to the ground state
of the system,  or if one performs a usual two-sublattice
DMFT calculation for the doped Hubbard model, the self-energy
oscillates during the self-consistency calculation and a converged
solution cannot be obtained.
We previously interpreted these oscillations as the tendency of the
system to form an SDW.\cite{Peters2009a,Peters2009b} However, this interpretation was 
mathematically not well founded and properties of the SDW state could not be
determined, because the DMFT calculation did not
converge. 

In this paper, we demonstrate how to overcome these above-mentioned
difficulties by performing large-scale simulations using
  the inhomogeneous DMFT (iDMFT) to 
study inhomogeneous phases in strongly correlated materials.
As an example, we study incommensurate SDW states in the doped
Hubbard model on a square lattice. However, the iDMFT is neither
restricted to the Hubbard model nor to the square lattice, but can be
employed for any strongly correlated model with local interactions.
We show that the 
oscillations, which have been observed in previous DMFT calculations,
indeed indicate the emergence of SDWs. 
The iDMFT is an extension of DMFT to incorporate
inhomogeneities and has been so far used to study surfaces,
interfaces, superlattices, and trapped strongly correlated
systems.\cite{Potthoff1999,Helmes2008,Snoek2008,Zenia2009,Gorelik2010,Snoek2011,Tada2012,Peters2013,Tada2013,Heikkinen2013,Peters2014}
Furthermore, there have been a few works in which the
iDMFT has been used for SDWs in the Hubbard model. However, these
calculations have been for small cluster sizes or one-dimensional
slices of the lattice. Thus, these calculations mainly focused on the
SDW state called vertical stripes in the context of the high
temperature superconducting cuprates.\cite{Fleck_2000,Fleck_2001,Raczkowski2010}
Using the iDMFT, we are
able to find a converged and self-consistent solution for the doped,
magnetically-ordered Hubbard model, and
thus are able to analyze different kinds of SDWs without a priori knowing
the rotation angle of the SDW.
Furthermore, the iDMFT gives us a direct access to dynamical
correlation functions, which is a great advantage over previous static
mean field calculations.

The remainder of this paper is organized as follows: In the next
section, we will 
explain technical details of the iDMFT calculations. 
This is followed by our results for the
SDW phase of the Hubbard model, including an analysis of static as well
as dynamical properties. A short summary will conclude this paper.

\section{technical details on the calculations}
\subsection{Dynamical mean field theory}
DMFT relates the lattice model to a quantum impurity model. This mapping
becomes exact in the limit of infinite dimensions. In this
high-dimensional limit, the hopping amplitude has to be scaled as
$t\rightarrow t^\star /\sqrt z$ ($z$: coordination number of the
lattice), in order to ensure a non-trivial kinetic energy. A
consequence of this scaling is the vanishing of the momentum
dependence of the self-energy, $\Sigma(k,\omega)\rightarrow \Sigma(\omega)$.

The local lattice Green's function can thus be written as 
\begin{eqnarray}
G_{loc}(z)&=&\frac{1}{N}\sum_k\frac{1}{z-\epsilon_k-\Sigma(z)}\nonumber\\
&=&\int d\epsilon \frac{\rho_0(\epsilon)}{z-\epsilon-\Sigma(z)},\label{locallatGreen}
\end{eqnarray}
where $\epsilon_k$ represents the noninteracting band structure of the
lattice, and $\rho(\epsilon)$ the corresponding noninteracting local
density of states (DOS).

The mapping onto a quantum impurity model can be done by comparing
the local lattice Green's function (Eq. (\ref{locallatGreen})) to the
Green's function of an impurity model with the same local interaction term,
which reads
\begin{displaymath}
G_{imp}(z)=\frac{1}{z-\Delta(z)-\Sigma(z)},
\end{displaymath}
where $\Delta(z)$ is the hybridization between the impurity level and
an electron bath.

An iteratively performed DMFT calculation is done as follows: First, with
a given self-energy, which can be zero in the first iteration, the
local lattice Green's function is calculated by
Eq. (\ref{locallatGreen}). Second, from this  local lattice Green's
function, one calculates the hybridization $\Delta(z)$ of the
corresponding quantum impurity model by
\begin{equation}
\Delta(z)=z-\left(G_{loc}(z)\right)^{-1}-\Sigma(z).\label{hybridization}
\end{equation}
This hybridization defines a quantum impurity model, whose self-energy
must be determined. With this self-energy, one calculates a new local
lattice Green's function from which the next
 quantum impurity model is determined. This procedure continues until a
converged solution is found.

The DMFT can also be used to investigate properties of long-range
ordered phases of strongly correlated models. When performing
calculations for a magnetic phase, one has to calculate a
spin-dependent self-energy, which results in a spin-dependent
hybridization of the quantum impurity model. In the case of an
antiferromagnetic N\'eel state,  one has to take into account the doubling
of the unit cell. The local lattice Green's function can then be
calculated by the so-called AB-sublattice method,
\begin{displaymath}
G_{loc}(z)=\int d\epsilon\rho_0(\epsilon) \left(\begin{array}{cc}z-\Sigma_\uparrow(z)&-\epsilon\\-\epsilon&z-\Sigma_\downarrow(z)\end{array}\right)^{-1}.
\end{displaymath}
As stated above, the AB-sublattice method works well for the
antiferromagnetic N\'eel 
state, where the spin direction alternates between nearest
neighbors. However, this method fails to describe long-range ordered
phases, which are not commensurate with two sublattices, e.g. the
antiferromagnetic state of the doped Hubbard model. In the next subsection,
we will show how to overcome this problem by using iDMFT.

\subsection{iDMFT for SDWs}
In order to stabilize a long-range ordered  SDW state with wavelength larger
than two lattice sites, one has to divide the lattice into large
enough clusters so that the wavelength of the ordered state can be
taken correctly into account. A way to do that is
to use the iDMFT, which maps each
lattice site of a cluster onto its corresponding quantum impurity model, thereby
assuming a momentum independent self-energy. Although the
self-energy between different lattice sites vanishes within this
approximation, the self-energy of each lattice site may be different.
The iDMFT can thus describe
inhomogeneous systems, such as cold atoms in a trap
potential, or interfaces and surfaces of strongly correlated
systems. We here apply the 
iDMFT for a homogeneous model, but in a situation where the symmetry
of the model is spontaneously broken, which results in
an inhomogeneous state.

The iDMFT works as follows: After setting the size of our cluster, usually
between $400$ and $2000$ lattice sites, we initialize a self-energy for each 
lattice site. In the first DMFT iteration, this self-energy can
be set to zero. We usually choose this self-energy in a
way that it breaks the SU(2) symmetry of the Hamiltonian in order to
obtain an SDW wave solution. (If the SU(2) symmetry is not broken, the
iDMFT solution will be a paramagnetic state.)
Using these self-energies, we calculate the
local Green's functions for all lattice sites by using a matrix inversion:
\begin{eqnarray}
\mathbf{G}_{loc}(z)=\left[z\cdot\mathbb{I}-\mathbf{H}-\mathbf{\Sigma} \right]^{-1},\label{ilocalGreen}
\end{eqnarray}
where $\mathbf{H}$ is the noninteracting Hamiltonian of the chosen cluster.
At this point, one must specify, if the calculation is performed for a
finite cluster with open or periodic boundary conditions, or if the
calculation is for an infinite lattice, which consists of repeating
this finite cluster. In the case of the infinite lattice, the
Hamiltonian will include momentum dependent terms, which must be
integrated over. Equation (\ref{ilocalGreen}) then reads
\begin{eqnarray*}
\mathbf{G}_{loc}(z)=\int dk_xdk_y\left(\left[z\cdot\mathbb{I}-\mathbf{H}_{k_x,k_y}-\mathbf{\Sigma} \right]^{-1}\right).
\end{eqnarray*}
After having calculated all local Green's functions, lattice-site dependent
quantum impurity models can be determined similar to
Eq. (\ref{hybridization}) by
\begin{equation}
\Delta_{ii}(z)=z-\left(G_{ii,loc}(z)\right)^{-1}-\Sigma_{ii}(z),
\end{equation}
(The inversion of the local Green's function is performed locally in order to determine the hybridization of a local
quantum impurity model.)
We now solve all these quantum impurity models and
calculate the corresponding self-energies. With these self-energies,
one can then 
calculate the local Green's function of the next iDMFT iteration, see
Eq. (\ref{ilocalGreen}).
Because the self-energy depends on the lattice site, 
this method is able to calculate properties of inhomogeneous phases
for this cluster.

During the iDMFT procedure, self-energies of several quantum impurity
models must be calculated. 
In this paper, we use the numerical renormalization
group (NRG),\cite{wilson1975,bulla2008} for this purpose. The NRG is a
well established method which is able to calculate numerical-exact
dynamical correlation functions such as Green's functions and
self-energies.\cite{peters2006,weichselbaum2007} 

These calculations thereby involve two time consuming
  steps: First, the calculation of the self energies of all lattice sites,
  which scales linearly with the number of lattice sites. The other time
  consuming step is the calculation of the local Green's functions,
  which involves a matrix inversion of the whole cluster and thus
  scales cubically with the number of lattice sites.

Although we are using a cluster of lattice sites in our calculations,
iDMFT is fundamentally different from the cluster DMFT (CDMFT) or the
dynamical cluster approximation (DCA).\cite{Maier2006} In iDMFT, one determines a local
self-energy separately for all lattice sites, while in the latter
methods, one solves a multi-site impurity model for the whole cluster,
thus determining also non-local terms of the self-energy. Therefore,
the CDMFT and the DCA are more accurate in principle, because they incorporate
non-local fluctuations. However, a cluster of approximately $1000$
sites is by far out of range for these methods.

\section{Doped antiferromagnetic state in the Hubbard model}
\begin{figure}[t]
\begin{center}
\includegraphics[width=1\linewidth]{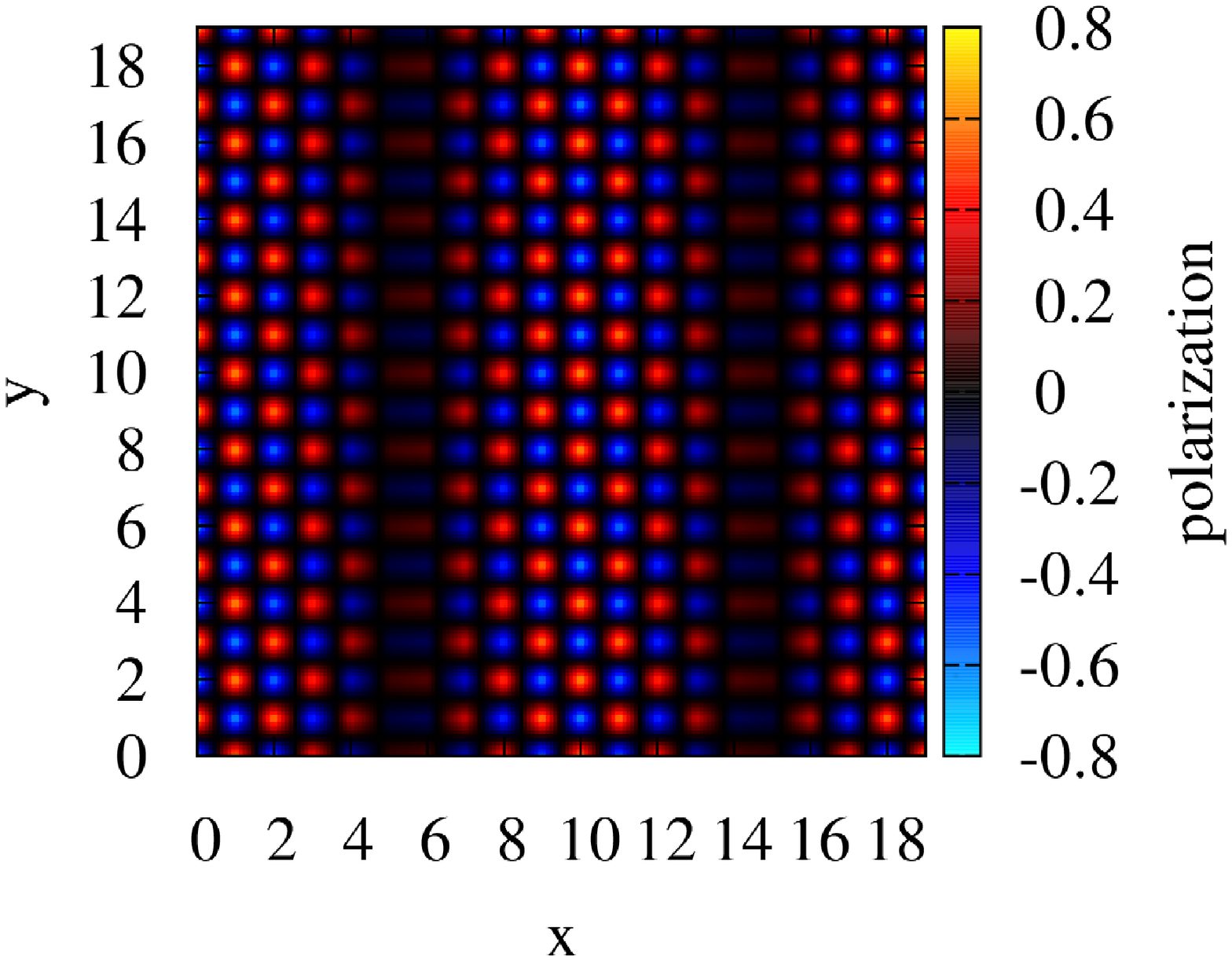}
\includegraphics[width=1\linewidth]{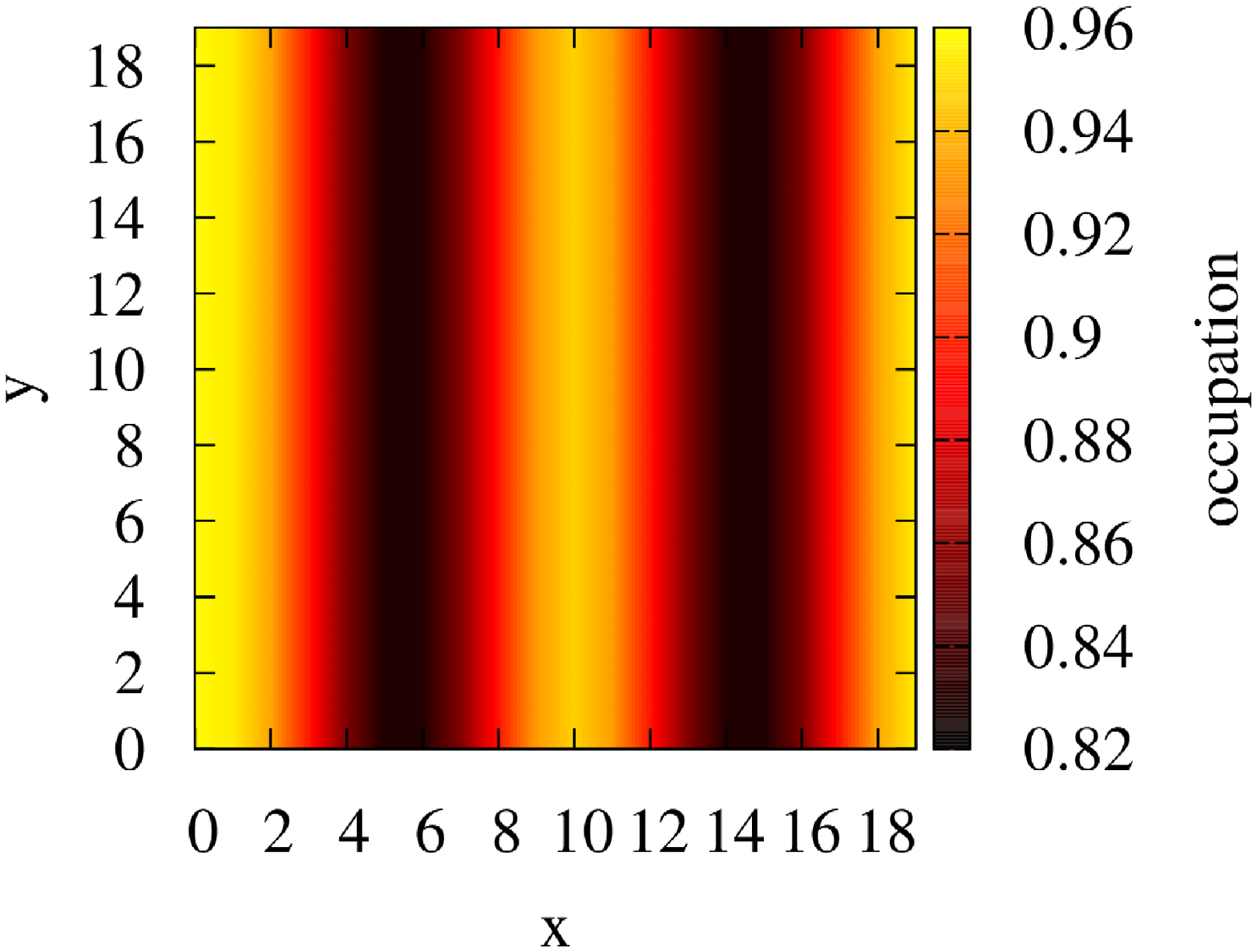}
\end{center}
\caption{(Color online) Typical pattern for a vertical SDW state in the Hubbard model
  for $U=8t$ and an average electron density $\langle
  n\rangle=0.9$. The upper (lower) panel shows the electron
  polarization (density).
 \label{SDW}}
\end{figure}
Intensive studies about incommensurate SDW states in the Hubbard model began
approximately at the same time as the discovery of high temperature
superconductivity\cite{Bednorz1986} that appears in strongly
correlated materials close 
to an antiferromagnetic
phase.\cite{Anderson_1987} These studies concerned the
antiferromagnetic phases in the $t$-$J$-model\cite{Shraiman_1989,
  Shraiman_1988, Kane_1990,
  Yoshioka_1989,Igarashi_1992,Igarashi_1992b,Jayaprakash_1989,Sarker_1991}
and the Hubbard model \cite{Schulz_1989,Schulz_1989b,poiblanc_1989,Zaanen_1989,Chu_1991,Yang_1991,Ichimura_1992,Chubukov_1995,Inui_1991,Dzierzawa_1992,Giamarchi_1990,Dombre_1990,Kato_1990,Dzierzawa_1993,Fresard_1991,Arrigoni_1991,Fleck_1998,Fleck_1999,Freericks_1995,Gora_1999,Xu_2011}
and mainly exploited different types of static mean field theory,
e.g. Hartree Fock theory.
Summarizing these results, one can say that an extended region of SDW
states exists in the phase diagram of
the Hubbard model close
to half filling. 
At weak coupling, these SDWs run along one axis in the $(0,1)$- or
$(1,0)$-direction, which are called vertical SDWs. For stronger
coupling, the energetically favored state is an 
SDW running along the diagonal of the square lattice.
Furthermore, the SDW state is accompanied by a charge density wave
in the same direction. 
For strong enough coupling, the doped holes
localize in straight lines, yielding large areas of nearly half-filled
antiferromagnetically ordered sites and paramagnetic 
stripes with particle number less than one. 
These states
are usually referred to as stripe- or domain-wall-states.
As already mentioned above, besides these static mean field
calculations, there have been up to now only a few 
DMFT calculations, because the two-sublattice method does not yield a
converged solution for the doped Hubbard model. 
There have been iDMFT calculations for small clusters, mainly
  1D cuts through the 2D lattice,\cite{Fleck_2000,Fleck_2001,Raczkowski2010} or DMFT calculations incorporating knowledge
about properties of the SDW,\cite{Fleck_1998,Fleck_1999} such as the ordering vector. Furthermore,
we want to note that there have been some density matrix
renormalization group-\cite{Bonca_2000,White_2003,Hager_2005,Fehske_2006,Machida_2009} and constraint-path quantum
Monte-Carlo-calculations\cite{Bonca_2000,Chang_2010} for stripes in the doped Hubbard
model. However, these simulations are also restricted to small cluster sizes.

In contrast to these previous approaches, we here use the iDMFT
for large clusters of lattice sites. This allows us to stabilize
different kinds of SDWs without knowledge about their properties,
such  as an ordering wave vector. Furthermore, the iDMFT incorporates local
fluctuations exactly and thus goes well beyond static mean field theory
and gives a direct access to dynamical properties, i.e. Green's functions and
self-energies.

\begin{figure}[t]
\begin{center}
\includegraphics[width=1\linewidth]{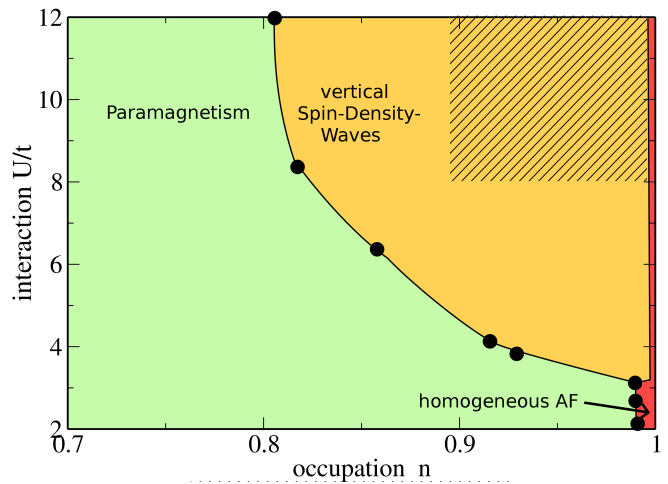}
\end{center}
\caption{(Color online) Phase diagram of the Hubbard on a square lattice as
  calculated by iDMFT. The shaded region represents parameters where
  we find vertical as well as diagonal SDWs to be stable. The
  homogeneous N\'eel state exists exactly at half filling for
  all interaction strengths and for a slightly doped region at weak interaction.
 \label{phase}}
\end{figure}
We have performed iDMFT calculations for finite cluster of at least
$400$ and up to $2000$ lattice sites using periodic boundary
conditions. In these calculations, we have found, 
in accordance with previous static mean field calculations, that doping
the antiferromagnetic N\'eel state in the Hubbard model results in SDW
states.
A typical solution of such an SDW state is shown in
Fig. \ref{SDW}. The upper (lower) panel shows the polarization
(occupation) of the lattice. In agreement with previous calculations, we see
that the SDW is accompanied by a charge density wave in the same
direction. 
The polarization of the electrons thereby depends on the
electron density. Regions of large electron density, which exhibit a
N\'eel-state-like order with large polarization, are separated by regions of
low electron density, which exhibit only a small polarization or are
magnetically disordered. For strong interaction, these regions of low
electron density form narrow straight lines, which have been
previously called stripes.
In Fig. \ref{SDW}, we observe that exactly at the center of these
stripes, there are always two neighboring sites which are 
ferromagnetically aligned. Thus, the N\'eel states of neighboring high
electron-density regions are phase-shifted. This is in accordance with
previous calculations.

In Fig. \ref{phase}, we summarize our calculations
in a phase diagram 
of the Hubbard model including SDW states.
For weak interaction strength, $U<3t$, we do not
observe any SDW phase. However, for these weak interactions,  the
antiferromagnetic  N\'eel state can be slightly doped without
destroying the  N\'eel order and exists up to $n\approx
0.97$ electrons per lattice. For $U>3t$, doping the N\'eel state
results in the emergence of  
SDW states in the Hubbard model. 
We have mainly focused on vertical SDW states,
which run along one of the axes of the square lattice,
see e.g. Fig. \ref{SDW}. These states 
have been identified by previous Hartree Fock calculations as the
ground state in the Hubbard model for moderate interaction strength.
In our calculations, these SDW states
are stable for strong enough interaction up to an electron
density $n\approx 0.8$. Compared to previous static mean field
calculations, our calculated 
critical occupation number is much closer to unity. This can be explained by
quantum fluctuations which are included in iDMFT but are absent in static mean
field calculations. The local interaction is screened by these quantum
 fluctuations. Thus, static mean field theories overestimate the
parameter region of ordered states.
We want to point out that besides the vertical 
SDW, different types of SDWs can be observed in iDMFT
calculations. 
In the whole parameter regime where vertical SDW states are stable,
SDW states which do not break the square lattice symmetry can also be
observed. These symmetric SDWs consist of modulations which run along 
both diagonals of the square lattice. 
Furthermore, for strong interaction, $U/t>7t$, diagonal SDW states which run
along a single diagonal of the square lattice can be observed close to half
filling. The shaded region in Fig. \ref{phase} corresponds to
parameters where we find diagonal SDW states to be stable.
These diagonal SDW states are unstable for weak interaction strength
and large doping. All stabilized SDW states are energetically very
close to each other; energetic differences are below our current accuracy.
However, compared to the paramagnetic state, any of these SDW
states is lower in energy.

Which state is realized in our iDMFT calculations, depends not only
on the energy of the state, but also on the way how
the symmetry is broken during the iDMFT calculation. If the SU(2) symmetry is
not broken at all, a paramagnetic state is formed. If the SU(2) symmetry
is broken at a single point, e.g. by applying a magnetic field in the
first iDMFT 
iteration at a single lattice site, then a square-lattice-symmetric state
arises. If a
magnetic field is applied to lattice sites in a vertical (diagonal) line, a
vertical (diagonal) SDW arises, if energetically stable.

\begin{figure}[t]
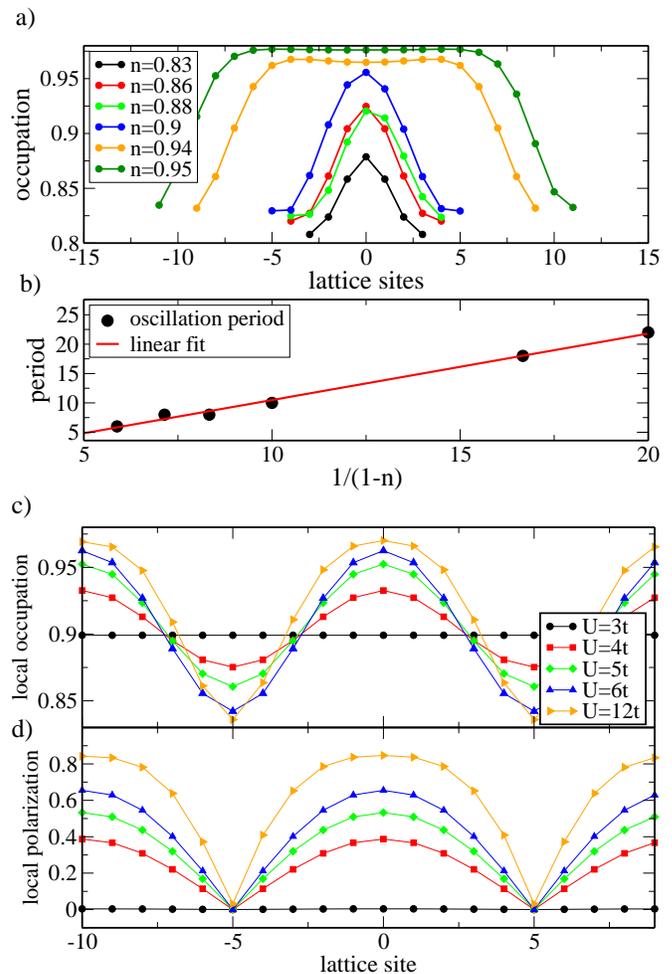

\begin{center}
\includegraphics[width=1\linewidth]{fig3a.eps}
\includegraphics[width=1\linewidth]{fig3b.eps}
\end{center}
\caption{(Color online) a) Electron density against lattice sites for
  different average electron densities and $U=8t$. We only show a single oscillation of
  the electron density which is periodically continued.
b) The extracted period of the modulation of the electron density against the
reciprocal density of holes, $1/(1-n)$ ($n$: average number of electrons per
lattice site) c) Electron density against lattice sites for
  different interaction strength and fixed average electron density
  $n=0.9$.
d) The magnitude of the spin polarization, $P=\vert
n_{\uparrow}-n_{\downarrow}\vert$,  against different lattice sites
for different interaction strength.
 \label{density_profile}}
\end{figure}

In order to present a more detailed analysis of SDW states,
we focus now on the vertical SDW state. Our results about vertical SDW states
are summarized in Fig. \ref{density_profile}. 
\begin{figure}[t]
\begin{center}
\includegraphics[width=1\linewidth]{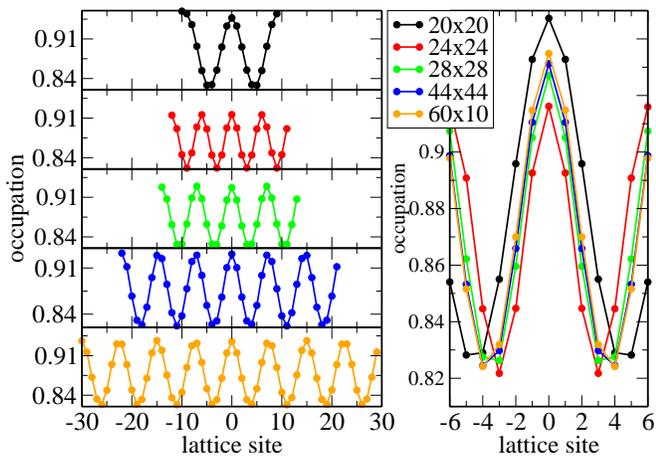}
\end{center}
\caption{(Color online) Left panels) Occupation number of different lattice sites of
  a vertical SDW for fixed chemical potential
  and interaction strength but different cluster sizes. Right panel)
  Direct comparison of a single period of the SDW for different
  cluster sizes. (The colors of the right panel correspond to the
  colors of the left panels).
 \label{finite_size_scaling}}
\end{figure}
In Fig. \ref{density_profile}  a) and b), we show the spatial modulation of the
electron density and the spin polarization for different occupation
numbers and interaction strengths.  
The period of the modulation
increases with increasing average electron density (the closer the average
occupation is to unity, the longer is the period). 
This is confirmed in panel b) of 
Fig. \ref{density_profile}, where we show that the period of
the modulation is approximately the inverse of the average number of
holes per lattice site, $p=1/(1-n)$, where $n$ is the average number
of electrons. 
Moreover, in panel a) it is visible that while for
low average electron densities 
the modulation is sine-like, the maximum is flattened, if the
occupation is close to unity; there are more and more half filled
lattice sites that exhibit a N\'eel state like order, while holes are
located within narrow  walls. These states have been called
domain-wall states or stripes in previous mean field calculations. 
While the period of the modulation strongly depends on the average electron
density, it is independent of the
interaction strength.
For increasing the interaction value for a fixed
occupation number (panel c) and d) in Fig. \ref{density_profile}), the
amplitude of the modulation becomes larger, and thus the SDW becomes more stripe-like.
The maximum of the local electron
density increases while the minimum decreases. At the same time, also the
maximum of the electron polarization increases. However, the modulation
period remains unchanged, if the average filling of the lattice is not
changed.
These static properties of SDW states, such as amplitude
and modulation period, qualitatively agree with previous static mean field
calculations. 

In contrast to previous attempts to use DMFT, we
are here using large-scale iDMFT calculations. We are thus able to
analyze the influence of finite size effects on our
calculations. Because of the usage of periodic
boundary conditions, an integer number of oscillations of the SDW must
be included within the cluster. If the cluster size does not match the
period of the 
energetically most stable SDW, which is related to the occupation
number of the system, two things may happen in the system: First, the period
of the SDW may be slightly modified. Second, while most oscillations of
the SDW within the cluster correspond to the most stable SDW, there
are a few oscillations which are altered in order to accommodate the
SDW within the cluster. Figure \ref{finite_size_scaling} shows the occupation
profiles of vertical SDWs for interaction strength $U=8t$ and chemical
potential $\mu=2t$ for different cluster sizes. The left panels show
the occupation number of different lattice sites of the SDW for
different cluster sizes. These panels show that the 
qualitative structure of the SDW does only weakly depend on the
cluster size. A direct comparison of a single oscillation of the SDWs
on these different clusters is presented in the right panel of
Fig. \ref{finite_size_scaling}. The average occupation for this
chemical potential is approximately $\langle n\rangle=0.875$
corresponding to an SDW with period $8$. We here show a comparison of
an oscillation in the middle of our cluster, where we have initially
broken the SU(2) symmetry and thus the SDW is usually best converged
within the cluster. We observe that while the SDWs for the clusters
$20\times 20$ and $24\times 24$ have different periods and different amplitudes of
the occupation, the SDWs for cluster sizes larger than $28$ and the
rectangular cluster look very similar. The size of the cluster which
is needed to obtain reliable results does of course depend on the
period of the SDW. We observe that the cluster size should be
approximately $4$ times the period of the SDW, which makes it more and
more difficult to calculate SDWs close to half filling with
very long periods.

\begin{figure}[t]
\begin{center}
\includegraphics[width=1\linewidth]{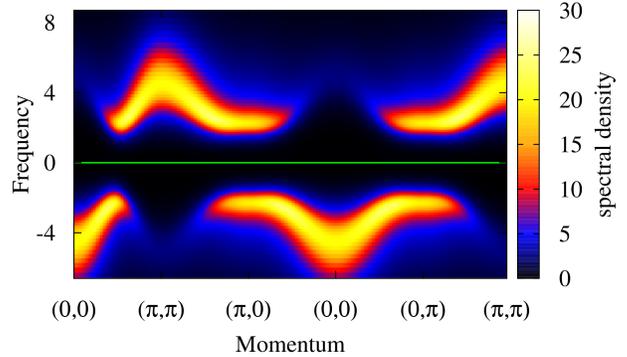}
\end{center}
\caption{(Color online) Momentum resolved spectral function for the half filled 
  Hubbard model for $U=8t$. The green line marks the Fermi energy. 
 \label{spec_half}}
\end{figure}
\begin{figure}[t]
\begin{center}
\includegraphics[width=1\linewidth]{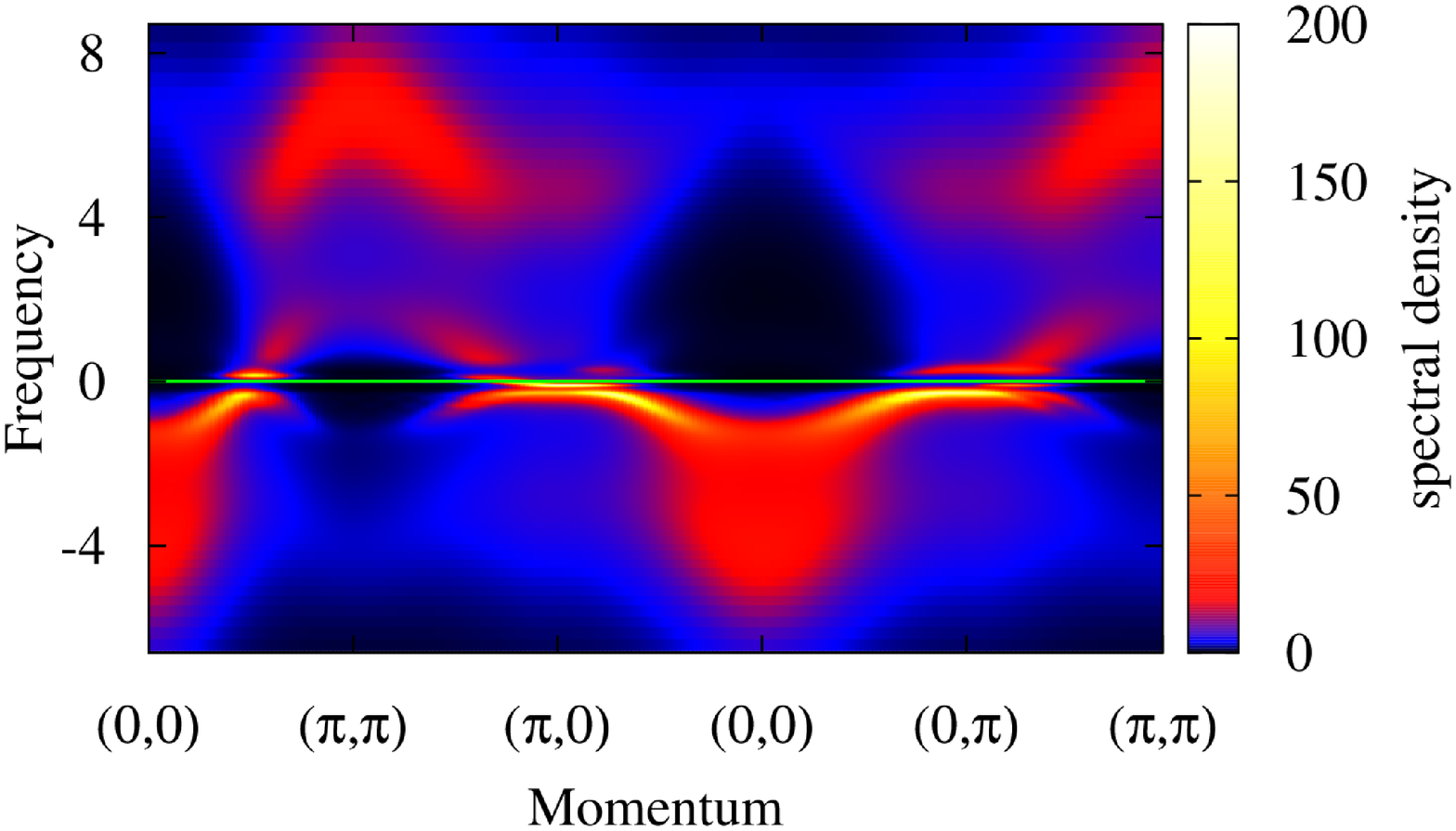}
\includegraphics[width=1\linewidth]{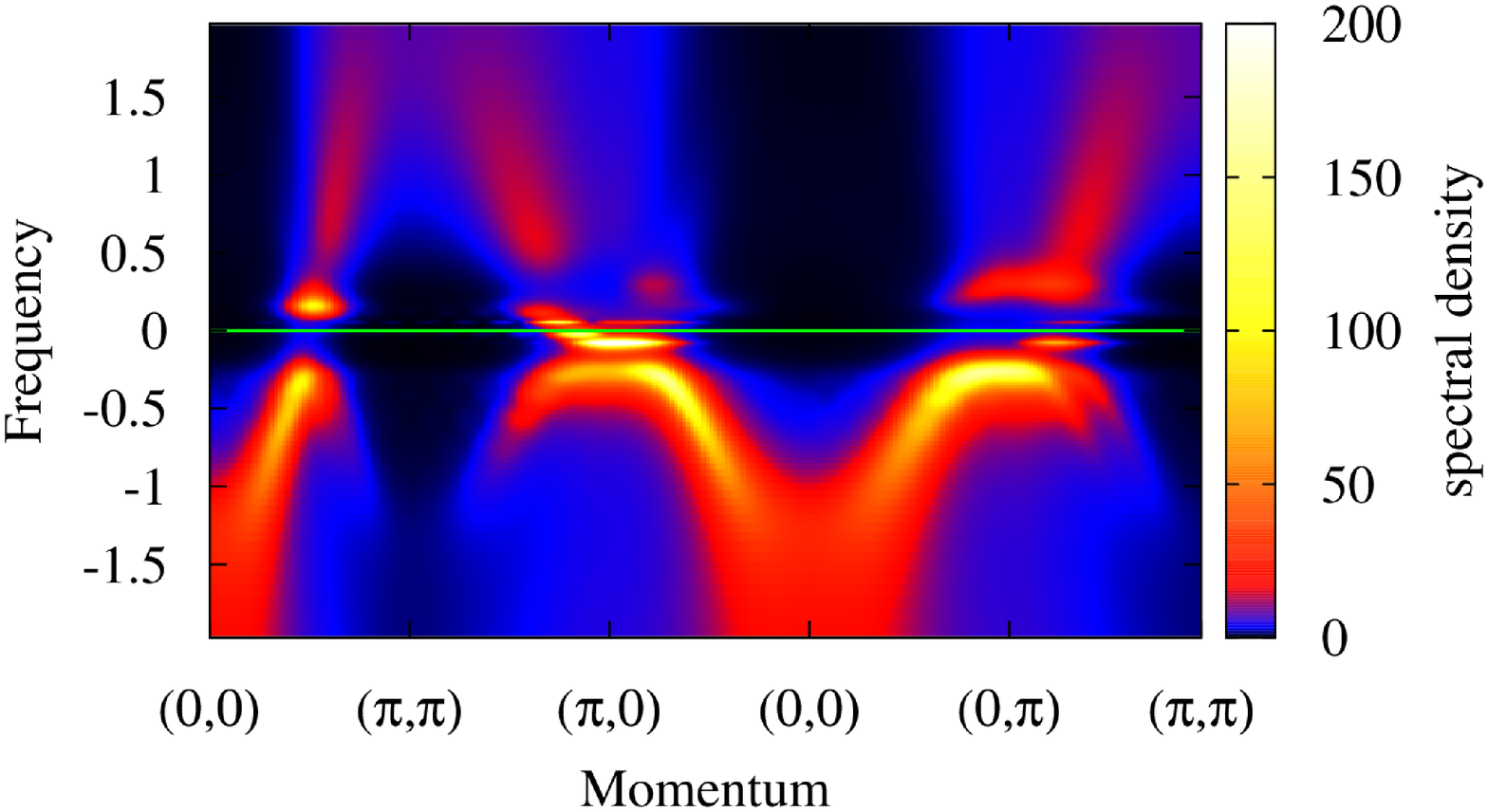}
\end{center}
\caption{(Color online) Momentum resolved spectral function for the doped
  Hubbard model for $U=8t$, $\langle n\rangle=0.95$ in the vertical
  SDW state. The green line
  marks the Fermi energy. The lower panel is a magnification around
  the Fermi energy.
 \label{spec_dope}}
\end{figure}
\begin{figure}[t]
\begin{center}
\includegraphics[width=1\linewidth]{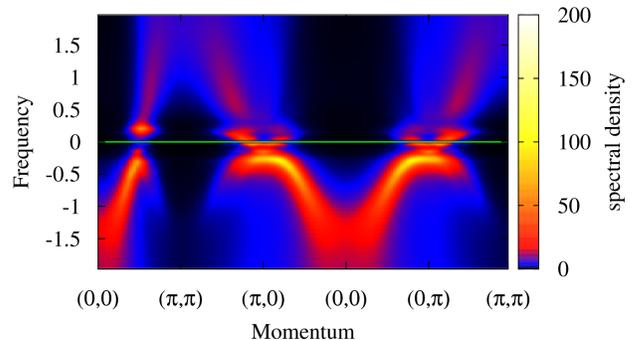}
\end{center}
\caption{(Color online) Momentum resolved spectral function for the doped
  Hubbard model for $U=8t$, $\langle n\rangle=0.95$ in the
  square-lattice symmetric SDW state.  The green line
  marks the Fermi energy.
 \label{spectral_point}}
\end{figure}

An advantage of iDMFT over static mean field calculations is the
easy access to dynamical properties; in fact, dynamical properties
such as
self-energies are calculated in order to perform an iDMFT
procedure. It should be noted that the Green's function takes two
lattice sites as indices, $G_{(x_1,y_1),(x_2,y_2)}$, because properties
vary among different lattice sites. Therefore, the Green's function
cannot be written in terms of the distance
between lattice sites.
However, in order to show a DOS which does
only depend on a single momentum $\vec k = (k_x,k_y)$, we Fourier
transform the distance of lattice sites, $\vec 
d=(x_1-x_2,y_1-y_2)$, and average it over the whole lattice. The resulting
momentum-resolved spectral function corresponds then to the DOS which
would be measured e.g. in angle-resolved-photo-emission spectroscopy. We thus calculate
\begin{eqnarray*}
G_{k_x,k_y}(\omega)&=&\frac{1}{N}\sum_{x_1,y_1}\sum_{x_2,y_2}\Bigl( G_{(x_1,y_1),(x_2,y_2)}(\omega)\times\\&&\exp(i(k_x(x_1-x_2)+k_y(y_1-y_2))\Bigr).
\end{eqnarray*}
The momentum resolved spectral functions shown below are then given by
$\rho_{k_x,k_y}(\omega)=-\frac{1}{\pi}\text{Im} (G_{k_x,k_y}(\omega))$.

The momentum resolved DOS of a half filled
Hubbard model for $U=8t$ is shown in Fig. \ref{spec_half}. 
The system is in an insulating antiferromagnetic N\'eel state. The DOS
exhibits a gap around the Fermi energy as denoted by the green
line. If we dope holes into the system, this state changes into a
vertical SDW state.
The
momentum-resolved DOS for a vertical SDW with  
average electron density $\langle n\rangle=0.95$ is shown in Fig. \ref{spec_dope}.
We observe that the lower band crosses the Fermi energy at three
points in momentum space: $\vec
k=(\pi/2,\pi/2)$, $\vec k=(0,\pi)$, and $\vec k=(\pi,0)$. 
In a magnification of the DOS around the Fermi energy (lower panel of
Fig. \ref{spec_dope}), however, we see that the band which crosses
the Fermi energy becomes gapped at $\vec k=(\pi/2,\pi/2)$ and $\vec
k=(0,\pi)$. There is a small but finite gap at the Fermi energy. The
lower band possesses spectral weight only for $\vec k=(\pi,0)$ at the
Fermi energy. This ``pseudo-gap'' originates in the SDW order. 
In the vertical SDW state, which is shown in Fig. \ref{spec_dope}, the
spectrum becomes gapped for $\vec k=(\pi/2,\pi/2)$ and one of the two
momenta  $\vec k=(\pi,0)$ or  $\vec k=(0,\pi)$ depending on the
direction of the vertical SDW. The momentum, where the spectrum at the
Fermi energy is not gapped, thereby corresponds
to the direction of the SDW; if there is spectral weight at the Fermi
energy for $\vec k=(\pi,0)$, then the SDW runs along the
$x$-directions. This means that there are paramagnetic stripes in the
$y$-direction. These paramagnetic stripes are the cause for the 
spectral weight at the momentum $\vec k=(\pi,0)$.
For comparison, we show the spectral function of the
square-lattice-symmetric SDW state in Fig. \ref{spectral_point}. Due
to symmetry, 
the spectral weight around $\vec k=(\pi,0)$ and $\vec k=(0,\pi)$ is
equal. Furthermore, both momenta possess spectral weight at the Fermi
energy. The spectrum only becomes gapped at the Fermi energy for $\vec
k=(\pi/2,\pi/2)$. We can therefore say that this pseudo gap in the
spectral function is clearly related to the SDW order.

\section{Conclusions}
We have used the iDMFT for the calculation of
incommensurate SDW states in strongly correlated electron systems such
as the Hubbard model and have thereby resolved difficulties
encountered in previous DMFT treatments of magnetic states away from
half filling.  We have calculated the magnetic phase diagram
of the Hubbard model on a square lattice including SDW states, and
have shown that screening of the Coulomb interaction due to local
fluctuations, which  cannot be taken into account in static mean
field 
calculations, strongly modifies the phase diagram. As a result,
magnetically ordered phases vanish for average electron densities
$n<0.8$.
 We have focused in this paper on
vertical SDWs, although different types of SDW can be
stabilized. The calculated properties, such as the period of the SDW,
agree very well with previous calculations. However, a great
advantage of iDMFT over static mean field calculations is the easy
access to dynamical properties such as momentum resolved spectral
functions. We have shown that due to the SDW order, parts of the
spectrum at the Fermi
energy become gapped. The Fermi momenta at which the spectrum is gapped
are directly related to the type of SDW state, e.g. in which direction
the vertical SDW runs.

Finally, we want to stress that this method is not limited to the
SDWs in the Hubbard on a square lattice, but can also be used for
studying properties of incommensurate ordered phases for
various 2D as well as 3D lattices. Furthermore, it can be easily
adopted to study different strongly correlated models as long as the
interaction is local.

%%%%%%%%%%%%%%%%%%%%%%%%%%
\begin{acknowledgments}
%%%%%%%%%%%%%%%%%%%%%%%%%%
RP and NK thank the Japan Society for the Promotion of Science (JSPS)
for the support  through its FIRST Program. RP thanks RIKEN for the
support through the FPR program.
NK acknowledges support through KAKENHI (No. 22103005, No. 25400366).
The numerical 
calculations were performed at the ISSP 
in Tokyo and on the SR16000 at  YITP in Kyoto University.
\end{acknowledgments}


\begin{thebibliography}{69}
\expandafter\ifx\csname natexlab\endcsname\relax\def\natexlab#1{#1}\fi
\expandafter\ifx\csname bibnamefont\endcsname\relax
  \def\bibnamefont#1{#1}\fi
\expandafter\ifx\csname bibfnamefont\endcsname\relax
  \def\bibfnamefont#1{#1}\fi
\expandafter\ifx\csname citenamefont\endcsname\relax
  \def\citenamefont#1{#1}\fi
\expandafter\ifx\csname url\endcsname\relax
  \def\url#1{\texttt{#1}}\fi
\expandafter\ifx\csname urlprefix\endcsname\relax\def\urlprefix{URL }\fi
\providecommand{\bibinfo}[2]{#2}
\providecommand{\eprint}[2][]{\url{#2}}

\bibitem[{\citenamefont{Hubbard}(1963)}]{hubbard1963}
\bibinfo{author}{\bibfnamefont{J.}~\bibnamefont{Hubbard}},
  \bibinfo{journal}{Proc. R. Soc. A} \textbf{\bibinfo{volume}{276}},
  \bibinfo{pages}{238} (\bibinfo{year}{1963}).

\bibitem[{\citenamefont{Kanamori}(1963)}]{kanamori1963}
\bibinfo{author}{\bibfnamefont{J.}~\bibnamefont{Kanamori}},
  \bibinfo{journal}{Prog. Theor. Phys.} \textbf{\bibinfo{volume}{30}},
  \bibinfo{pages}{275} (\bibinfo{year}{1963}).

\bibitem[{\citenamefont{Gutzwiller}(1963)}]{gutzwiller1963}
\bibinfo{author}{\bibfnamefont{M.~C.} \bibnamefont{Gutzwiller}},
  \bibinfo{journal}{Phys. Rev. Lett.} \textbf{\bibinfo{volume}{10}},
  \bibinfo{pages}{159} (\bibinfo{year}{1963}).

\bibitem[{\citenamefont{Imada et~al.}(1998)\citenamefont{Imada, Fujimori, and
  Tokura}}]{imada1998}
\bibinfo{author}{\bibfnamefont{M.}~\bibnamefont{Imada}},
  \bibinfo{author}{\bibfnamefont{A.}~\bibnamefont{Fujimori}}, \bibnamefont{and}
  \bibinfo{author}{\bibfnamefont{Y.}~\bibnamefont{Tokura}},
  \bibinfo{journal}{Rev. Mod. Phys.} \textbf{\bibinfo{volume}{70}},
  \bibinfo{pages}{1039} (\bibinfo{year}{1998}).

\bibitem[{\citenamefont{Metzner and Vollhardt}(1989)}]{Metzner1989}
\bibinfo{author}{\bibfnamefont{W.}~\bibnamefont{Metzner}} \bibnamefont{and}
  \bibinfo{author}{\bibfnamefont{D.}~\bibnamefont{Vollhardt}},
  \bibinfo{journal}{Phys. Rev. Lett.} \textbf{\bibinfo{volume}{62}},
  \bibinfo{pages}{324} (\bibinfo{year}{1989}),
  \urlprefix\url{http://link.aps.org/doi/10.1103/PhysRevLett.62.324}.

\bibitem[{\citenamefont{Pruschke et~al.}(1995)\citenamefont{Pruschke, Jarrell,
  and Freericks}}]{Pruschke1995}
\bibinfo{author}{\bibfnamefont{T.}~\bibnamefont{Pruschke}},
  \bibinfo{author}{\bibfnamefont{M.}~\bibnamefont{Jarrell}}, \bibnamefont{and}
  \bibinfo{author}{\bibfnamefont{J.}~\bibnamefont{Freericks}},
  \bibinfo{journal}{Adv.~Phys.} \textbf{\bibinfo{volume}{44}},
  \bibinfo{pages}{187} (\bibinfo{year}{1995}).

\bibitem[{\citenamefont{Georges et~al.}(1996)\citenamefont{Georges, Kotliar,
  Krauth, and Rozenberg}}]{Georges1996}
\bibinfo{author}{\bibfnamefont{A.}~\bibnamefont{Georges}},
  \bibinfo{author}{\bibfnamefont{G.}~\bibnamefont{Kotliar}},
  \bibinfo{author}{\bibfnamefont{W.}~\bibnamefont{Krauth}}, \bibnamefont{and}
  \bibinfo{author}{\bibfnamefont{M.~J.} \bibnamefont{Rozenberg}},
  \bibinfo{journal}{Rev. Mod. Phys.} \textbf{\bibinfo{volume}{68}},
  \bibinfo{pages}{13} (\bibinfo{year}{1996}),
  \urlprefix\url{http://link.aps.org/doi/10.1103/RevModPhys.68.13}.

\bibitem[{\citenamefont{Maier et~al.}(2005)\citenamefont{Maier, Jarrell,
  Pruschke, and Hettler}}]{Maier2006}
\bibinfo{author}{\bibfnamefont{T.}~\bibnamefont{Maier}},
  \bibinfo{author}{\bibfnamefont{M.}~\bibnamefont{Jarrell}},
  \bibinfo{author}{\bibfnamefont{T.}~\bibnamefont{Pruschke}}, \bibnamefont{and}
  \bibinfo{author}{\bibfnamefont{M.~H.} \bibnamefont{Hettler}},
  \bibinfo{journal}{Rev. Mod. Phys.} \textbf{\bibinfo{volume}{77}},
  \bibinfo{pages}{1027} (\bibinfo{year}{2005}),
  \urlprefix\url{http://link.aps.org/doi/10.1103/RevModPhys.77.1027}.

\bibitem[{\citenamefont{Jarrell}(1992)}]{Jarrell1992}
\bibinfo{author}{\bibfnamefont{M.}~\bibnamefont{Jarrell}},
  \bibinfo{journal}{Phys. Rev. Lett.} \textbf{\bibinfo{volume}{69}},
  \bibinfo{pages}{168} (\bibinfo{year}{1992}),
  \urlprefix\url{http://link.aps.org/doi/10.1103/PhysRevLett.69.168}.

\bibitem[{\citenamefont{Freericks and Jarrell}(1995)}]{Freericks_1995}
\bibinfo{author}{\bibfnamefont{J.~K.} \bibnamefont{Freericks}}
  \bibnamefont{and} \bibinfo{author}{\bibfnamefont{M.}~\bibnamefont{Jarrell}},
  \bibinfo{journal}{Phys. Rev. Lett.} \textbf{\bibinfo{volume}{74}},
  \bibinfo{pages}{186} (\bibinfo{year}{1995}),
  \urlprefix\url{http://link.aps.org/doi/10.1103/PhysRevLett.74.186}.

\bibitem[{\citenamefont{van Dongen}(1995)}]{dongen1995}
\bibinfo{author}{\bibfnamefont{P.~G.~J.} \bibnamefont{van Dongen}},
  \bibinfo{journal}{Phys. Rev. Lett.} \textbf{\bibinfo{volume}{74}},
  \bibinfo{pages}{182} (\bibinfo{year}{1995}),
  \urlprefix\url{http://link.aps.org/doi/10.1103/PhysRevLett.74.182}.

\bibitem[{\citenamefont{van Dongen}(1996)}]{dongen1996}
\bibinfo{author}{\bibfnamefont{P.~G.~J.} \bibnamefont{van Dongen}},
  \bibinfo{journal}{Phys. Rev. B} \textbf{\bibinfo{volume}{54}},
  \bibinfo{pages}{1584} (\bibinfo{year}{1996}),
  \urlprefix\url{http://link.aps.org/doi/10.1103/PhysRevB.54.1584}.

\bibitem[{\citenamefont{Zitzler et~al.}(2002)\citenamefont{Zitzler, Pruschke,
  and Bulla}}]{zitzler2002}
\bibinfo{author}{\bibfnamefont{R.}~\bibnamefont{Zitzler}},
  \bibinfo{author}{\bibfnamefont{T.}~\bibnamefont{Pruschke}}, \bibnamefont{and}
  \bibinfo{author}{\bibfnamefont{R.}~\bibnamefont{Bulla}},
  \bibinfo{journal}{The European Physical Journal B - Condensed Matter and
  Complex Systems} \textbf{\bibinfo{volume}{27}}, \bibinfo{pages}{473}
  (\bibinfo{year}{2002}), ISSN \bibinfo{issn}{1434-6028},
  \urlprefix\url{http://dx.doi.org/10.1140/epjb/e2002-00180-3}.

\bibitem[{\citenamefont{Peters and Pruschke}(2009{\natexlab{a}})}]{Peters2009a}
\bibinfo{author}{\bibfnamefont{R.}~\bibnamefont{Peters}} \bibnamefont{and}
  \bibinfo{author}{\bibfnamefont{T.}~\bibnamefont{Pruschke}},
  \bibinfo{journal}{Phys. Rev. B} \textbf{\bibinfo{volume}{79}},
  \bibinfo{pages}{045108} (\bibinfo{year}{2009}{\natexlab{a}}),
  \urlprefix\url{http://link.aps.org/doi/10.1103/PhysRevB.79.045108}.

\bibitem[{\citenamefont{Peters and Pruschke}(2009{\natexlab{b}})}]{Peters2009b}
\bibinfo{author}{\bibfnamefont{R.}~\bibnamefont{Peters}} \bibnamefont{and}
  \bibinfo{author}{\bibfnamefont{T.}~\bibnamefont{Pruschke}},
  \bibinfo{journal}{New Journal of Physics} \textbf{\bibinfo{volume}{11}},
  \bibinfo{pages}{083022} (\bibinfo{year}{2009}{\natexlab{b}}),
  \urlprefix\url{http://stacks.iop.org/1367-2630/11/i=8/a=083022}.

\bibitem[{\citenamefont{Fleck et~al.}(1998)\citenamefont{Fleck, Liechtenstein,
  Ole\ifmmode~\acute{s}\else \'{s}\fi{}, Hedin, and Anisimov}}]{Fleck_1998}
\bibinfo{author}{\bibfnamefont{M.}~\bibnamefont{Fleck}},
  \bibinfo{author}{\bibfnamefont{A.~I.} \bibnamefont{Liechtenstein}},
  \bibinfo{author}{\bibfnamefont{A.~M.} \bibnamefont{Ole\ifmmode~\acute{s}\else
  \'{s}\fi{}}}, \bibinfo{author}{\bibfnamefont{L.}~\bibnamefont{Hedin}},
  \bibnamefont{and} \bibinfo{author}{\bibfnamefont{V.~I.}
  \bibnamefont{Anisimov}}, \bibinfo{journal}{Phys. Rev. Lett.}
  \textbf{\bibinfo{volume}{80}}, \bibinfo{pages}{2393} (\bibinfo{year}{1998}),
  \urlprefix\url{http://link.aps.org/doi/10.1103/PhysRevLett.80.2393}.

\bibitem[{\citenamefont{Fleck et~al.}(1999)\citenamefont{Fleck, Lichtenstein,
  Ole\ifmmode~\acute{s}\else \'{s}\fi{}, and Hedin}}]{Fleck_1999}
\bibinfo{author}{\bibfnamefont{M.}~\bibnamefont{Fleck}},
  \bibinfo{author}{\bibfnamefont{A.~I.} \bibnamefont{Lichtenstein}},
  \bibinfo{author}{\bibfnamefont{A.~M.} \bibnamefont{Ole\ifmmode~\acute{s}\else
  \'{s}\fi{}}}, \bibnamefont{and}
  \bibinfo{author}{\bibfnamefont{L.}~\bibnamefont{Hedin}},
  \bibinfo{journal}{Phys. Rev. B} \textbf{\bibinfo{volume}{60}},
  \bibinfo{pages}{5224} (\bibinfo{year}{1999}),
  \urlprefix\url{http://link.aps.org/doi/10.1103/PhysRevB.60.5224}.

\bibitem[{\citenamefont{Potthoff and Nolting}(1999)}]{Potthoff1999}
\bibinfo{author}{\bibfnamefont{M.}~\bibnamefont{Potthoff}} \bibnamefont{and}
  \bibinfo{author}{\bibfnamefont{W.}~\bibnamefont{Nolting}},
  \bibinfo{journal}{Phys. Rev. B} \textbf{\bibinfo{volume}{59}},
  \bibinfo{pages}{2549} (\bibinfo{year}{1999}),
  \urlprefix\url{http://link.aps.org/doi/10.1103/PhysRevB.59.2549}.

\bibitem[{\citenamefont{Helmes et~al.}(2008)\citenamefont{Helmes, Costi, and
  Rosch}}]{Helmes2008}
\bibinfo{author}{\bibfnamefont{R.~W.} \bibnamefont{Helmes}},
  \bibinfo{author}{\bibfnamefont{T.~A.} \bibnamefont{Costi}}, \bibnamefont{and}
  \bibinfo{author}{\bibfnamefont{A.}~\bibnamefont{Rosch}},
  \bibinfo{journal}{Phys. Rev. Lett.} \textbf{\bibinfo{volume}{101}},
  \bibinfo{pages}{066802} (\bibinfo{year}{2008}),
  \urlprefix\url{http://link.aps.org/doi/10.1103/PhysRevLett.101.066802}.

\bibitem[{\citenamefont{Snoek et~al.}(2008)\citenamefont{Snoek, Titvinidze,
  T\"oke, Byczuk, and Hofstetter}}]{Snoek2008}
\bibinfo{author}{\bibfnamefont{M.}~\bibnamefont{Snoek}},
  \bibinfo{author}{\bibfnamefont{I.}~\bibnamefont{Titvinidze}},
  \bibinfo{author}{\bibfnamefont{C.}~\bibnamefont{T\"oke}},
  \bibinfo{author}{\bibfnamefont{K.}~\bibnamefont{Byczuk}}, \bibnamefont{and}
  \bibinfo{author}{\bibfnamefont{W.}~\bibnamefont{Hofstetter}},
  \bibinfo{journal}{New Journal of Physics} \textbf{\bibinfo{volume}{10}},
  \bibinfo{pages}{093008} (\bibinfo{year}{2008}),
  \urlprefix\url{http://stacks.iop.org/1367-2630/10/i=9/a=093008}.

\bibitem[{\citenamefont{Zenia et~al.}(2009)\citenamefont{Zenia, Freericks,
  Krishnamurthy, and Pruschke}}]{Zenia2009}
\bibinfo{author}{\bibfnamefont{H.}~\bibnamefont{Zenia}},
  \bibinfo{author}{\bibfnamefont{J.~K.} \bibnamefont{Freericks}},
  \bibinfo{author}{\bibfnamefont{H.~R.} \bibnamefont{Krishnamurthy}},
  \bibnamefont{and} \bibinfo{author}{\bibfnamefont{T.}~\bibnamefont{Pruschke}},
  \bibinfo{journal}{Phys. Rev. Lett.} \textbf{\bibinfo{volume}{103}},
  \bibinfo{pages}{116402} (\bibinfo{year}{2009}),
  \urlprefix\url{http://link.aps.org/doi/10.1103/PhysRevLett.103.116402}.

\bibitem[{\citenamefont{Gorelik et~al.}(2010)\citenamefont{Gorelik, Titvinidze,
  Hofstetter, Snoek, and Bl\"umer}}]{Gorelik2010}
\bibinfo{author}{\bibfnamefont{E.~V.} \bibnamefont{Gorelik}},
  \bibinfo{author}{\bibfnamefont{I.}~\bibnamefont{Titvinidze}},
  \bibinfo{author}{\bibfnamefont{W.}~\bibnamefont{Hofstetter}},
  \bibinfo{author}{\bibfnamefont{M.}~\bibnamefont{Snoek}}, \bibnamefont{and}
  \bibinfo{author}{\bibfnamefont{N.}~\bibnamefont{Bl\"umer}},
  \bibinfo{journal}{Phys. Rev. Lett.} \textbf{\bibinfo{volume}{105}},
  \bibinfo{pages}{065301} (\bibinfo{year}{2010}),
  \urlprefix\url{http://link.aps.org/doi/10.1103/PhysRevLett.105.065301}.

\bibitem[{\citenamefont{Snoek et~al.}(2011)\citenamefont{Snoek, Titvinidze, and
  Hofstetter}}]{Snoek2011}
\bibinfo{author}{\bibfnamefont{M.}~\bibnamefont{Snoek}},
  \bibinfo{author}{\bibfnamefont{I.}~\bibnamefont{Titvinidze}},
  \bibnamefont{and}
  \bibinfo{author}{\bibfnamefont{W.}~\bibnamefont{Hofstetter}},
  \bibinfo{journal}{Phys. Rev. B} \textbf{\bibinfo{volume}{83}},
  \bibinfo{pages}{054419} (\bibinfo{year}{2011}),
  \urlprefix\url{http://link.aps.org/doi/10.1103/PhysRevB.83.054419}.

\bibitem[{\citenamefont{Tada et~al.}(2012)\citenamefont{Tada, Peters, Oshikawa,
  Koga, Kawakami, and Fujimoto}}]{Tada2012}
\bibinfo{author}{\bibfnamefont{Y.}~\bibnamefont{Tada}},
  \bibinfo{author}{\bibfnamefont{R.}~\bibnamefont{Peters}},
  \bibinfo{author}{\bibfnamefont{M.}~\bibnamefont{Oshikawa}},
  \bibinfo{author}{\bibfnamefont{A.}~\bibnamefont{Koga}},
  \bibinfo{author}{\bibfnamefont{N.}~\bibnamefont{Kawakami}}, \bibnamefont{and}
  \bibinfo{author}{\bibfnamefont{S.}~\bibnamefont{Fujimoto}},
  \bibinfo{journal}{Phys. Rev. B} \textbf{\bibinfo{volume}{85}},
  \bibinfo{pages}{165138} (\bibinfo{year}{2012}),
  \urlprefix\url{http://link.aps.org/doi/10.1103/PhysRevB.85.165138}.

\bibitem[{\citenamefont{Peters et~al.}(2013)\citenamefont{Peters, Tada, and
  Kawakami}}]{Peters2013}
\bibinfo{author}{\bibfnamefont{R.}~\bibnamefont{Peters}},
  \bibinfo{author}{\bibfnamefont{Y.}~\bibnamefont{Tada}}, \bibnamefont{and}
  \bibinfo{author}{\bibfnamefont{N.}~\bibnamefont{Kawakami}},
  \bibinfo{journal}{Phys. Rev. B} \textbf{\bibinfo{volume}{88}},
  \bibinfo{pages}{155134} (\bibinfo{year}{2013}),
  \urlprefix\url{http://link.aps.org/doi/10.1103/PhysRevB.88.155134}.

\bibitem[{\citenamefont{Tada et~al.}(2013)\citenamefont{Tada, Peters, and
  Oshikawa}}]{Tada2013}
\bibinfo{author}{\bibfnamefont{Y.}~\bibnamefont{Tada}},
  \bibinfo{author}{\bibfnamefont{R.}~\bibnamefont{Peters}}, \bibnamefont{and}
  \bibinfo{author}{\bibfnamefont{M.}~\bibnamefont{Oshikawa}},
  \bibinfo{journal}{Phys. Rev. B} \textbf{\bibinfo{volume}{88}},
  \bibinfo{pages}{235121} (\bibinfo{year}{2013}),
  \urlprefix\url{http://link.aps.org/doi/10.1103/PhysRevB.88.235121}.

\bibitem[{\citenamefont{Heikkinen et~al.}(2013)\citenamefont{Heikkinen, Kim,
  and T\"orm\"a}}]{Heikkinen2013}
\bibinfo{author}{\bibfnamefont{M.~O.~J.} \bibnamefont{Heikkinen}},
  \bibinfo{author}{\bibfnamefont{D.-H.} \bibnamefont{Kim}}, \bibnamefont{and}
  \bibinfo{author}{\bibfnamefont{P.}~\bibnamefont{T\"orm\"a}},
  \bibinfo{journal}{Phys. Rev. B} \textbf{\bibinfo{volume}{87}},
  \bibinfo{pages}{224513} (\bibinfo{year}{2013}),
  \urlprefix\url{http://link.aps.org/doi/10.1103/PhysRevB.87.224513}.

\bibitem[{\citenamefont{Peters and Kawakami}(2014)}]{Peters2014}
\bibinfo{author}{\bibfnamefont{R.}~\bibnamefont{Peters}} \bibnamefont{and}
  \bibinfo{author}{\bibfnamefont{N.}~\bibnamefont{Kawakami}},
  \bibinfo{journal}{Phys. Rev. B} \textbf{\bibinfo{volume}{89}},
  \bibinfo{pages}{041106} (\bibinfo{year}{2014}),
  \urlprefix\url{http://link.aps.org/doi/10.1103/PhysRevB.89.041106}.

\bibitem[{\citenamefont{Fleck et~al.}(2000)\citenamefont{Fleck, Lichtenstein,
  Pavarini, and Ole\ifmmode~\acute{s}\else \'{s}\fi{}}}]{Fleck_2000}
\bibinfo{author}{\bibfnamefont{M.}~\bibnamefont{Fleck}},
  \bibinfo{author}{\bibfnamefont{A.~I.} \bibnamefont{Lichtenstein}},
  \bibinfo{author}{\bibfnamefont{E.}~\bibnamefont{Pavarini}}, \bibnamefont{and}
  \bibinfo{author}{\bibfnamefont{A.~M.} \bibnamefont{Ole\ifmmode~\acute{s}\else
  \'{s}\fi{}}}, \bibinfo{journal}{Phys. Rev. Lett.}
  \textbf{\bibinfo{volume}{84}}, \bibinfo{pages}{4962} (\bibinfo{year}{2000}),
  \urlprefix\url{http://link.aps.org/doi/10.1103/PhysRevLett.84.4962}.

\bibitem[{\citenamefont{Fleck et~al.}(2001)\citenamefont{Fleck, Lichtenstein,
  and Ole\ifmmode~\acute{s}\else \'{s}\fi{}}}]{Fleck_2001}
\bibinfo{author}{\bibfnamefont{M.}~\bibnamefont{Fleck}},
  \bibinfo{author}{\bibfnamefont{A.~I.} \bibnamefont{Lichtenstein}},
  \bibnamefont{and} \bibinfo{author}{\bibfnamefont{A.~M.}
  \bibnamefont{Ole\ifmmode~\acute{s}\else \'{s}\fi{}}}, \bibinfo{journal}{Phys.
  Rev. B} \textbf{\bibinfo{volume}{64}}, \bibinfo{pages}{134528}
  (\bibinfo{year}{2001}),
  \urlprefix\url{http://link.aps.org/doi/10.1103/PhysRevB.64.134528}.

\bibitem[{\citenamefont{Raczkowski and Assaad}(2010)}]{Raczkowski2010}
\bibinfo{author}{\bibfnamefont{M.}~\bibnamefont{Raczkowski}} \bibnamefont{and}
  \bibinfo{author}{\bibfnamefont{F.~F.} \bibnamefont{Assaad}},
  \bibinfo{journal}{Phys. Rev. B} \textbf{\bibinfo{volume}{82}},
  \bibinfo{pages}{233101} (\bibinfo{year}{2010}),
  \urlprefix\url{http://link.aps.org/doi/10.1103/PhysRevB.82.233101}.

\bibitem[{\citenamefont{Wilson}(1975)}]{wilson1975}
\bibinfo{author}{\bibfnamefont{K.}~\bibnamefont{Wilson}},
  \bibinfo{journal}{Rev. Mod. Phys.} \textbf{\bibinfo{volume}{47}},
  \bibinfo{pages}{773} (\bibinfo{year}{1975}).

\bibitem[{\citenamefont{Bulla et~al.}(2008)\citenamefont{Bulla, Costi, and
  Pruschke}}]{bulla2008}
\bibinfo{author}{\bibfnamefont{R.}~\bibnamefont{Bulla}},
  \bibinfo{author}{\bibfnamefont{T.}~\bibnamefont{Costi}}, \bibnamefont{and}
  \bibinfo{author}{\bibfnamefont{T.}~\bibnamefont{Pruschke}},
  \bibinfo{journal}{Rev. Mod. Phys.} \textbf{\bibinfo{volume}{80}},
  \bibinfo{pages}{395} (\bibinfo{year}{2008}).

\bibitem[{\citenamefont{Peters et~al.}(2006)\citenamefont{Peters, Pruschke, and
  Anders}}]{peters2006}
\bibinfo{author}{\bibfnamefont{R.}~\bibnamefont{Peters}},
  \bibinfo{author}{\bibfnamefont{T.}~\bibnamefont{Pruschke}}, \bibnamefont{and}
  \bibinfo{author}{\bibfnamefont{F.}~\bibnamefont{Anders}},
  \bibinfo{journal}{Phys. Rev. B} \textbf{\bibinfo{volume}{74}},
  \bibinfo{pages}{245114} (\bibinfo{year}{2006}).

\bibitem[{\citenamefont{Weichselbaum and von Delft}(2007)}]{weichselbaum2007}
\bibinfo{author}{\bibfnamefont{A.}~\bibnamefont{Weichselbaum}}
  \bibnamefont{and} \bibinfo{author}{\bibfnamefont{J.}~\bibnamefont{von
  Delft}}, \bibinfo{journal}{Phys. Rev. Lett.} \textbf{\bibinfo{volume}{99}},
  \bibinfo{pages}{076402} (\bibinfo{year}{2007}).

\bibitem[{\citenamefont{Bednorz and Müller}(1986)}]{Bednorz1986}
\bibinfo{author}{\bibfnamefont{J.}~\bibnamefont{Bednorz}} \bibnamefont{and}
  \bibinfo{author}{\bibfnamefont{K.}~\bibnamefont{Müller}},
  \bibinfo{journal}{Zeitschrift für Physik B Condensed Matter}
  \textbf{\bibinfo{volume}{64}}, \bibinfo{pages}{189} (\bibinfo{year}{1986}),
  ISSN \bibinfo{issn}{0722-3277},
  \urlprefix\url{http://dx.doi.org/10.1007/BF01303701}.

\bibitem[{\citenamefont{Anderson}(1987)}]{Anderson_1987}
\bibinfo{author}{\bibfnamefont{P.~W.} \bibnamefont{Anderson}},
  \bibinfo{journal}{Science} \textbf{\bibinfo{volume}{235}},
  \bibinfo{pages}{1196} (\bibinfo{year}{1987}),
  \eprint{http://www.sciencemag.org/content/235/4793/1196.full.pdf},
  \urlprefix\url{http://www.sciencemag.org/content/235/4793/1196.abstract}.

\bibitem[{\citenamefont{Shraiman and Siggia}(1989)}]{Shraiman_1989}
\bibinfo{author}{\bibfnamefont{B.~I.} \bibnamefont{Shraiman}} \bibnamefont{and}
  \bibinfo{author}{\bibfnamefont{E.~D.} \bibnamefont{Siggia}},
  \bibinfo{journal}{Phys. Rev. Lett.} \textbf{\bibinfo{volume}{62}},
  \bibinfo{pages}{1564} (\bibinfo{year}{1989}),
  \urlprefix\url{http://link.aps.org/doi/10.1103/PhysRevLett.62.1564}.

\bibitem[{\citenamefont{Shraiman and Siggia}(1988)}]{Shraiman_1988}
\bibinfo{author}{\bibfnamefont{B.~I.} \bibnamefont{Shraiman}} \bibnamefont{and}
  \bibinfo{author}{\bibfnamefont{E.~D.} \bibnamefont{Siggia}},
  \bibinfo{journal}{Phys. Rev. Lett.} \textbf{\bibinfo{volume}{61}},
  \bibinfo{pages}{467} (\bibinfo{year}{1988}),
  \urlprefix\url{http://link.aps.org/doi/10.1103/PhysRevLett.61.467}.

\bibitem[{\citenamefont{Kane et~al.}(1990)\citenamefont{Kane, Lee, Ng,
  Chakraborty, and Read}}]{Kane_1990}
\bibinfo{author}{\bibfnamefont{C.~L.} \bibnamefont{Kane}},
  \bibinfo{author}{\bibfnamefont{P.~A.} \bibnamefont{Lee}},
  \bibinfo{author}{\bibfnamefont{T.~K.} \bibnamefont{Ng}},
  \bibinfo{author}{\bibfnamefont{B.}~\bibnamefont{Chakraborty}},
  \bibnamefont{and} \bibinfo{author}{\bibfnamefont{N.}~\bibnamefont{Read}},
  \bibinfo{journal}{Phys. Rev. B} \textbf{\bibinfo{volume}{41}},
  \bibinfo{pages}{2653} (\bibinfo{year}{1990}),
  \urlprefix\url{http://link.aps.org/doi/10.1103/PhysRevB.41.2653}.

\bibitem[{\citenamefont{Yoshioka}(1989)}]{Yoshioka_1989}
\bibinfo{author}{\bibfnamefont{D.}~\bibnamefont{Yoshioka}},
  \bibinfo{journal}{Journal of the Physical Society of Japan}
  \textbf{\bibinfo{volume}{58}}, \bibinfo{pages}{1516} (\bibinfo{year}{1989}),
  \urlprefix\url{http://jpsj.ipap.jp/link?JPSJ/58/1516/}.

\bibitem[{\citenamefont{Igarashi and
  Fulde}(1992{\natexlab{a}})}]{Igarashi_1992}
\bibinfo{author}{\bibfnamefont{J.-i.} \bibnamefont{Igarashi}} \bibnamefont{and}
  \bibinfo{author}{\bibfnamefont{P.}~\bibnamefont{Fulde}},
  \bibinfo{journal}{Phys. Rev. B} \textbf{\bibinfo{volume}{45}},
  \bibinfo{pages}{10419} (\bibinfo{year}{1992}{\natexlab{a}}),
  \urlprefix\url{http://link.aps.org/doi/10.1103/PhysRevB.45.10419}.

\bibitem[{\citenamefont{Igarashi and
  Fulde}(1992{\natexlab{b}})}]{Igarashi_1992b}
\bibinfo{author}{\bibfnamefont{J.-i.} \bibnamefont{Igarashi}} \bibnamefont{and}
  \bibinfo{author}{\bibfnamefont{P.}~\bibnamefont{Fulde}},
  \bibinfo{journal}{Phys. Rev. B} \textbf{\bibinfo{volume}{45}},
  \bibinfo{pages}{12357} (\bibinfo{year}{1992}{\natexlab{b}}),
  \urlprefix\url{http://link.aps.org/doi/10.1103/PhysRevB.45.12357}.

\bibitem[{\citenamefont{Jayaprakash et~al.}(1989)\citenamefont{Jayaprakash,
  Krishnamurthy, and Sarker}}]{Jayaprakash_1989}
\bibinfo{author}{\bibfnamefont{C.}~\bibnamefont{Jayaprakash}},
  \bibinfo{author}{\bibfnamefont{H.~R.} \bibnamefont{Krishnamurthy}},
  \bibnamefont{and} \bibinfo{author}{\bibfnamefont{S.}~\bibnamefont{Sarker}},
  \bibinfo{journal}{Phys. Rev. B} \textbf{\bibinfo{volume}{40}},
  \bibinfo{pages}{2610} (\bibinfo{year}{1989}),
  \urlprefix\url{http://link.aps.org/doi/10.1103/PhysRevB.40.2610}.

\bibitem[{\citenamefont{Sarker et~al.}(1991)\citenamefont{Sarker, Jayaprakash,
  Krishnamurthy, and Wenzel}}]{Sarker_1991}
\bibinfo{author}{\bibfnamefont{S.}~\bibnamefont{Sarker}},
  \bibinfo{author}{\bibfnamefont{C.}~\bibnamefont{Jayaprakash}},
  \bibinfo{author}{\bibfnamefont{H.~R.} \bibnamefont{Krishnamurthy}},
  \bibnamefont{and} \bibinfo{author}{\bibfnamefont{W.}~\bibnamefont{Wenzel}},
  \bibinfo{journal}{Phys. Rev. B} \textbf{\bibinfo{volume}{43}},
  \bibinfo{pages}{8775} (\bibinfo{year}{1991}),
  \urlprefix\url{http://link.aps.org/doi/10.1103/PhysRevB.43.8775}.

\bibitem[{\citenamefont{Schulz}(1990)}]{Schulz_1989}
\bibinfo{author}{\bibfnamefont{H.~J.} \bibnamefont{Schulz}},
  \bibinfo{journal}{Phys. Rev. Lett.} \textbf{\bibinfo{volume}{64}},
  \bibinfo{pages}{1445} (\bibinfo{year}{1990}),
  \urlprefix\url{http://link.aps.org/doi/10.1103/PhysRevLett.64.1445}.

\bibitem[{\citenamefont{Schulz}(1989)}]{Schulz_1989b}
\bibinfo{author}{\bibfnamefont{H.~J.} \bibnamefont{Schulz}},
  \bibinfo{journal}{J. Phys. France} \textbf{\bibinfo{volume}{50}},
  \bibinfo{pages}{2833} (\bibinfo{year}{1989}),
  \urlprefix\url{http://dx.doi.org/10.1051/jphys:0198900500180283300}.

\bibitem[{\citenamefont{Poilblanc and Rice}(1989)}]{poiblanc_1989}
\bibinfo{author}{\bibfnamefont{D.}~\bibnamefont{Poilblanc}} \bibnamefont{and}
  \bibinfo{author}{\bibfnamefont{T.~M.} \bibnamefont{Rice}},
  \bibinfo{journal}{Phys. Rev. B} \textbf{\bibinfo{volume}{39}},
  \bibinfo{pages}{9749} (\bibinfo{year}{1989}),
  \urlprefix\url{http://link.aps.org/doi/10.1103/PhysRevB.39.9749}.

\bibitem[{\citenamefont{Zaanen and Gunnarsson}(1989)}]{Zaanen_1989}
\bibinfo{author}{\bibfnamefont{J.}~\bibnamefont{Zaanen}} \bibnamefont{and}
  \bibinfo{author}{\bibfnamefont{O.}~\bibnamefont{Gunnarsson}},
  \bibinfo{journal}{Phys. Rev. B} \textbf{\bibinfo{volume}{40}},
  \bibinfo{pages}{7391} (\bibinfo{year}{1989}),
  \urlprefix\url{http://link.aps.org/doi/10.1103/PhysRevB.40.7391}.

\bibitem[{\citenamefont{Chu}(1991)}]{Chu_1991}
\bibinfo{author}{\bibfnamefont{H.}~\bibnamefont{Chu}}, \bibinfo{journal}{Solid
  State Communications} \textbf{\bibinfo{volume}{80}}, \bibinfo{pages}{1003 }
  (\bibinfo{year}{1991}), ISSN \bibinfo{issn}{0038-1098},
  \urlprefix\url{http://www.sciencedirect.com/science/article/pii/003810989190408N}.

\bibitem[{\citenamefont{Yang and Su}(1991)}]{Yang_1991}
\bibinfo{author}{\bibfnamefont{J.}~\bibnamefont{Yang}} \bibnamefont{and}
  \bibinfo{author}{\bibfnamefont{W.~P.} \bibnamefont{Su}},
  \bibinfo{journal}{Phys. Rev. B} \textbf{\bibinfo{volume}{44}},
  \bibinfo{pages}{6838} (\bibinfo{year}{1991}),
  \urlprefix\url{http://link.aps.org/doi/10.1103/PhysRevB.44.6838}.

\bibitem[{\citenamefont{Ichimura et~al.}(1992)\citenamefont{Ichimura, Fujita,
  and Nakao}}]{Ichimura_1992}
\bibinfo{author}{\bibfnamefont{M.}~\bibnamefont{Ichimura}},
  \bibinfo{author}{\bibfnamefont{M.}~\bibnamefont{Fujita}}, \bibnamefont{and}
  \bibinfo{author}{\bibfnamefont{K.}~\bibnamefont{Nakao}},
  \bibinfo{journal}{Journal of the Physical Society of Japan}
  \textbf{\bibinfo{volume}{61}}, \bibinfo{pages}{2027} (\bibinfo{year}{1992}),
  \urlprefix\url{http://jpsj.ipap.jp/link?JPSJ/61/2027/}.

\bibitem[{\citenamefont{Chubukov and Musaelian}(1995)}]{Chubukov_1995}
\bibinfo{author}{\bibfnamefont{A.~V.} \bibnamefont{Chubukov}} \bibnamefont{and}
  \bibinfo{author}{\bibfnamefont{K.~A.} \bibnamefont{Musaelian}},
  \bibinfo{journal}{Phys. Rev. B} \textbf{\bibinfo{volume}{51}},
  \bibinfo{pages}{12605} (\bibinfo{year}{1995}),
  \urlprefix\url{http://link.aps.org/doi/10.1103/PhysRevB.51.12605}.

\bibitem[{\citenamefont{Inui and Littlewood}(1991)}]{Inui_1991}
\bibinfo{author}{\bibfnamefont{M.}~\bibnamefont{Inui}} \bibnamefont{and}
  \bibinfo{author}{\bibfnamefont{P.~B.} \bibnamefont{Littlewood}},
  \bibinfo{journal}{Phys. Rev. B} \textbf{\bibinfo{volume}{44}},
  \bibinfo{pages}{4415} (\bibinfo{year}{1991}),
  \urlprefix\url{http://link.aps.org/doi/10.1103/PhysRevB.44.4415}.

\bibitem[{\citenamefont{Dzierzawa}(1992)}]{Dzierzawa_1992}
\bibinfo{author}{\bibfnamefont{M.}~\bibnamefont{Dzierzawa}},
  \bibinfo{journal}{Zeitschrift für Physik B Condensed Matter}
  \textbf{\bibinfo{volume}{86}}, \bibinfo{pages}{49} (\bibinfo{year}{1992}),
  ISSN \bibinfo{issn}{0722-3277},
  \urlprefix\url{http://dx.doi.org/10.1007/BF01323546}.

\bibitem[{\citenamefont{Giamarchi and Lhuillier}(1990)}]{Giamarchi_1990}
\bibinfo{author}{\bibfnamefont{T.}~\bibnamefont{Giamarchi}} \bibnamefont{and}
  \bibinfo{author}{\bibfnamefont{C.}~\bibnamefont{Lhuillier}},
  \bibinfo{journal}{Phys. Rev. B} \textbf{\bibinfo{volume}{42}},
  \bibinfo{pages}{10641} (\bibinfo{year}{1990}),
  \urlprefix\url{http://link.aps.org/doi/10.1103/PhysRevB.42.10641}.

\bibitem[{\citenamefont{Dombre}(1990)}]{Dombre_1990}
\bibinfo{author}{\bibfnamefont{T.}~\bibnamefont{Dombre}}, \bibinfo{journal}{J.
  Phys. France} \textbf{\bibinfo{volume}{51}}, \bibinfo{pages}{847}
  (\bibinfo{year}{1990}),
  \urlprefix\url{http://dx.doi.org/10.1051/jphys:01990005109084700}.

\bibitem[{\citenamefont{Kato et~al.}(1990)\citenamefont{Kato, Machida,
  Nakanishi, and Fujita}}]{Kato_1990}
\bibinfo{author}{\bibfnamefont{M.}~\bibnamefont{Kato}},
  \bibinfo{author}{\bibfnamefont{K.}~\bibnamefont{Machida}},
  \bibinfo{author}{\bibfnamefont{H.}~\bibnamefont{Nakanishi}},
  \bibnamefont{and} \bibinfo{author}{\bibfnamefont{M.}~\bibnamefont{Fujita}},
  \bibinfo{journal}{Journal of the Physical Society of Japan}
  \textbf{\bibinfo{volume}{59}}, \bibinfo{pages}{1047} (\bibinfo{year}{1990}),
  \urlprefix\url{http://jpsj.ipap.jp/link?JPSJ/59/1047/}.

\bibitem[{\citenamefont{Dzierzawa and Frésard}(1993)}]{Dzierzawa_1993}
\bibinfo{author}{\bibfnamefont{M.}~\bibnamefont{Dzierzawa}} \bibnamefont{and}
  \bibinfo{author}{\bibfnamefont{R.}~\bibnamefont{Frésard}},
  \bibinfo{journal}{Zeitschrift für Physik B Condensed Matter}
  \textbf{\bibinfo{volume}{91}}, \bibinfo{pages}{245} (\bibinfo{year}{1993}),
  ISSN \bibinfo{issn}{0722-3277},
  \urlprefix\url{http://dx.doi.org/10.1007/BF01315242}.

\bibitem[{\citenamefont{Frésard et~al.}(1991)\citenamefont{Frésard,
  Dzierzawa, and Wölfle}}]{Fresard_1991}
\bibinfo{author}{\bibfnamefont{R.}~\bibnamefont{Frésard}},
  \bibinfo{author}{\bibfnamefont{M.}~\bibnamefont{Dzierzawa}},
  \bibnamefont{and} \bibinfo{author}{\bibfnamefont{P.}~\bibnamefont{Wölfle}},
  \bibinfo{journal}{EPL (Europhysics Letters)} \textbf{\bibinfo{volume}{15}},
  \bibinfo{pages}{325} (\bibinfo{year}{1991}),
  \urlprefix\url{http://stacks.iop.org/0295-5075/15/i=3/a=016}.

\bibitem[{\citenamefont{Arrigoni and Strinati}(1991)}]{Arrigoni_1991}
\bibinfo{author}{\bibfnamefont{E.}~\bibnamefont{Arrigoni}} \bibnamefont{and}
  \bibinfo{author}{\bibfnamefont{G.~C.} \bibnamefont{Strinati}},
  \bibinfo{journal}{Phys. Rev. B} \textbf{\bibinfo{volume}{44}},
  \bibinfo{pages}{7455} (\bibinfo{year}{1991}),
  \urlprefix\url{http://link.aps.org/doi/10.1103/PhysRevB.44.7455}.

\bibitem[{\citenamefont{G\'ora et~al.}(1999)\citenamefont{G\'ora,
  Ro\ifmmode~\acute{s}\else \'{s}\fi{}ciszewski, and Ole\ifmmode~\acute{s}\else
  \'{s}\fi{}}}]{Gora_1999}
\bibinfo{author}{\bibfnamefont{D.}~\bibnamefont{G\'ora}},
  \bibinfo{author}{\bibfnamefont{K.}~\bibnamefont{Ro\ifmmode~\acute{s}\else
  \'{s}\fi{}ciszewski}}, \bibnamefont{and}
  \bibinfo{author}{\bibfnamefont{A.~M.} \bibnamefont{Ole\ifmmode~\acute{s}\else
  \'{s}\fi{}}}, \bibinfo{journal}{Phys. Rev. B} \textbf{\bibinfo{volume}{60}},
  \bibinfo{pages}{7429} (\bibinfo{year}{1999}),
  \urlprefix\url{http://link.aps.org/doi/10.1103/PhysRevB.60.7429}.

\bibitem[{\citenamefont{Xu et~al.}(2011)\citenamefont{Xu, Chang, Walter, and
  Zhang}}]{Xu_2011}
\bibinfo{author}{\bibfnamefont{J.}~\bibnamefont{Xu}},
  \bibinfo{author}{\bibfnamefont{C.-C.} \bibnamefont{Chang}},
  \bibinfo{author}{\bibfnamefont{E.~J.} \bibnamefont{Walter}},
  \bibnamefont{and} \bibinfo{author}{\bibfnamefont{S.}~\bibnamefont{Zhang}},
  \bibinfo{journal}{Journal of Physics: Condensed Matter}
  \textbf{\bibinfo{volume}{23}}, \bibinfo{pages}{505601}
  (\bibinfo{year}{2011}),
  \urlprefix\url{http://stacks.iop.org/0953-8984/23/i=50/a=505601}.

\bibitem[{\citenamefont{Bon\ifmmode~\check{c}\else \v{c}\fi{}a
  et~al.}(2000)\citenamefont{Bon\ifmmode~\check{c}\else \v{c}\fi{}a,
  Gubernatis, Guerrero, Jeckelmann, and White}}]{Bonca_2000}
\bibinfo{author}{\bibfnamefont{J.}~\bibnamefont{Bon\ifmmode~\check{c}\else
  \v{c}\fi{}a}}, \bibinfo{author}{\bibfnamefont{J.~E.}
  \bibnamefont{Gubernatis}},
  \bibinfo{author}{\bibfnamefont{M.}~\bibnamefont{Guerrero}},
  \bibinfo{author}{\bibfnamefont{E.}~\bibnamefont{Jeckelmann}},
  \bibnamefont{and} \bibinfo{author}{\bibfnamefont{S.~R.} \bibnamefont{White}},
  \bibinfo{journal}{Phys. Rev. B} \textbf{\bibinfo{volume}{61}},
  \bibinfo{pages}{3251} (\bibinfo{year}{2000}),
  \urlprefix\url{http://link.aps.org/doi/10.1103/PhysRevB.61.3251}.

\bibitem[{\citenamefont{White and Scalapino}(2003)}]{White_2003}
\bibinfo{author}{\bibfnamefont{S.~R.} \bibnamefont{White}} \bibnamefont{and}
  \bibinfo{author}{\bibfnamefont{D.~J.} \bibnamefont{Scalapino}},
  \bibinfo{journal}{Phys. Rev. Lett.} \textbf{\bibinfo{volume}{91}},
  \bibinfo{pages}{136403} (\bibinfo{year}{2003}),
  \urlprefix\url{http://link.aps.org/doi/10.1103/PhysRevLett.91.136403}.

\bibitem[{\citenamefont{Hager et~al.}(2005)\citenamefont{Hager, Wellein,
  Jeckelmann, and Fehske}}]{Hager_2005}
\bibinfo{author}{\bibfnamefont{G.}~\bibnamefont{Hager}},
  \bibinfo{author}{\bibfnamefont{G.}~\bibnamefont{Wellein}},
  \bibinfo{author}{\bibfnamefont{E.}~\bibnamefont{Jeckelmann}},
  \bibnamefont{and} \bibinfo{author}{\bibfnamefont{H.}~\bibnamefont{Fehske}},
  in \emph{\bibinfo{booktitle}{High Performance Computing in Science and
  Engineering, Munich 2004}}, edited by
  \bibinfo{editor}{\bibfnamefont{S.}~\bibnamefont{Wagner}},
  \bibinfo{editor}{\bibfnamefont{W.}~\bibnamefont{Hanke}},
  \bibinfo{editor}{\bibfnamefont{A.}~\bibnamefont{Bode}}, \bibnamefont{and}
  \bibinfo{editor}{\bibfnamefont{F.}~\bibnamefont{Durst}}
  (\bibinfo{publisher}{Springer Berlin Heidelberg}, \bibinfo{year}{2005}), pp.
  \bibinfo{pages}{339--347}, ISBN \bibinfo{isbn}{978-3-540-44326-1},
  \urlprefix\url{http://dx.doi.org/10.1007/3-540-26657-7_31}.

\bibitem[{\citenamefont{Fehske et~al.}(2006)\citenamefont{Fehske, Hager,
  Wellein, and Jeckelmann}}]{Fehske_2006}
\bibinfo{author}{\bibfnamefont{H.}~\bibnamefont{Fehske}},
  \bibinfo{author}{\bibfnamefont{G.}~\bibnamefont{Hager}},
  \bibinfo{author}{\bibfnamefont{G.}~\bibnamefont{Wellein}}, \bibnamefont{and}
  \bibinfo{author}{\bibfnamefont{E.}~\bibnamefont{Jeckelmann}},
  \bibinfo{journal}{Physica B: Condensed Matter}
  \textbf{\bibinfo{volume}{378–380}}, \bibinfo{pages}{319 }
  (\bibinfo{year}{2006}), ISSN \bibinfo{issn}{0921-4526},
  \bibinfo{note}{proceedings of the International Conference on Strongly
  Correlated Electron Systems \{SCES\} 2005 Proceedings of the International
  Conference on Strongly Correlated Electron Systems},
  \urlprefix\url{http://www.sciencedirect.com/science/article/pii/S0921452606001530}.

\bibitem[{\citenamefont{Machida et~al.}(2009)\citenamefont{Machida, Okumura,
  Yamada, Ohashi, and Matsumoto}}]{Machida_2009}
\bibinfo{author}{\bibfnamefont{M.}~\bibnamefont{Machida}},
  \bibinfo{author}{\bibfnamefont{M.}~\bibnamefont{Okumura}},
  \bibinfo{author}{\bibfnamefont{S.}~\bibnamefont{Yamada}},
  \bibinfo{author}{\bibfnamefont{Y.}~\bibnamefont{Ohashi}}, \bibnamefont{and}
  \bibinfo{author}{\bibfnamefont{H.}~\bibnamefont{Matsumoto}},
  \bibinfo{journal}{Journal of Superconductivity and Novel Magnetism}
  \textbf{\bibinfo{volume}{22}}, \bibinfo{pages}{275} (\bibinfo{year}{2009}),
  ISSN \bibinfo{issn}{1557-1939},
  \urlprefix\url{http://dx.doi.org/10.1007/s10948-008-0438-5}.

\bibitem[{\citenamefont{Chang and Zhang}(2010)}]{Chang_2010}
\bibinfo{author}{\bibfnamefont{C.-C.} \bibnamefont{Chang}} \bibnamefont{and}
  \bibinfo{author}{\bibfnamefont{S.}~\bibnamefont{Zhang}},
  \bibinfo{journal}{Phys. Rev. Lett.} \textbf{\bibinfo{volume}{104}},
  \bibinfo{pages}{116402} (\bibinfo{year}{2010}),
  \urlprefix\url{http://link.aps.org/doi/10.1103/PhysRevLett.104.116402}.

\end{thebibliography}
\end{document}